\documentclass{icrc29}
\usepackage{graphicx,amssymb,amsmath,times}
\begin{document}
\title[``Dark Particle Accelerators''..]
{On the Probable Nature of the ``Dark Particle Accelerators'' Discovered by HESS}
\author[Abhas Mitra] {Abhas Mitra
        \newauthor\\
         Nuclear Research Lab., Bhabha Atomic Research Centre, 
            Mumbai 400 085, India\\        }
\presenter{Presenter: Abhas Mitra (amitra@apsara.barc.ernet.in), \  
ind-mitra-A-abs2-og22-poster}

\maketitle

\begin{abstract}
We estimate that the vacuum electric field associated with a spinning
Magnetospheric Eternally Collapsing Object MECO\cite{1,2,3,4,5,6,7,8,9} could
be {\em higher by a factor of} $\sim 10^4$ than the corresponding pulsar
value because of extreme relativistic Frame Dragging Effect. Thus isolated
spinning MECOs could be source of UHE cosmic ray acceleration and VHE $\gamma$-ray
production. However because of the steeply varying gravitational field close
to the surface of the MECO, any signal generated there would be both extremely redshifted and distorted. As a result, there may not be
any significant pulsed X-ray or Radio emission from close to the surface, and
consequently, the $\gamma$-ray source may appear as ``Unidentified'' and the
particle accelerator may appear as ``Dark''\cite{10,11}.

\end{abstract}

\section{Introduction}
The  VHE $\gamma$-ray telescope, 
HESS, has unrivalled sensitivity and success.
It is  in the process of revolutionizing the dormant and sluggish existing
 VHE $\gamma$-ray catalog the same way EGRET revolutionized GeV $\gamma$-ray catalog
almost 10 years ago.
 Accordingly it has been discovering ``unidentified'' sources
of VHE $\gamma$-ray sources\cite{10,11}. While some of such unidentified sources already discovered
or those which will be discovered routinely, could turn out to be sources about
which we already have idea, we predict that,  some of the sources, could
belong to an entirely new class.

For galactic sources, in the former category, we can have (1) Radio Pulsars,
(2) Pulsar Wind Nebulae (PWNe) (3) Supernova Remnants (SNRs), (4) X-ray binaries
containing Neutron Stars (NS), (5) Magnetars, (6) X-ray binaries containg supposed 
Black Holes Candidates (BHCs), with Microquasars being a sub-class,
(7) Hot Wolf -Rayet Star Wind Shocks\cite{12} (8) Hot O-B Star Wind Shocks\cite{11}.

In the latter category, we may have (9) Magnetospheric Eternally Collapsing Objects
(MECOs) as the entirely new class of VHE $\gamma$- ray sources. MECOs are end products
of collapse of massive stars  and held in quasistatic (unstable) equilibrium by
internal Eddington limited radiation luminosity 
and sperstrong (local) magnetic pressure\cite{1,2,3,4,5,6,7,8,9}. Eternally Collapsing 
Objects (ECOs) have also been briefly discussed in
ref.\cite{13}.
The surface redshift of a spherical object with radius $R$ is
\begin{equation}
z = (1-2GM/Rc^2)^{-1/2}  -1 = (1-r_g/R)^{-1/2} -1;\qquad r_g= 2GM/c^2 =30 (M/10 M_\odot)
Km
\end{equation}
where $G$ is the gravitational constant and $c$ is the speed of light. During gravitational
collapse of massive stars, as $z$
tends to increase indefinitely, the internal radiation and heat get
 virtually trapped, and,  catastrophic
collapse is halted. It is then that a MECO is formed. It is only when one overlooks this
general relativistic possibility of {\em virtual trapping of internal radiation}
 (which has been done
for past 90 years), one would have ``trapped surfaces'' ``Event Horizons'' and
(finite mass) BHs. 
While degeneracy pressure supported {\em cold \& strictly static} compact objects must have
$z <2$ and $M < (3-4) M_\odot$, ECOs/MECOs being {\em hot \& unstable},
 may have arbitrary high but finite $z$ and $M$.
In particular, Robertson \& Leiter\cite{2,3,4} have shown that for the MECOs lying in the X-ray
binaries, one expects, $z \sim 10^{7-8}$ and magnetic moment, {\em as seen by a distant
observer}, $\mu \sim 10^{30}$ Gauss cm$^3$.  While this latter value of $\mu$ is comparable
with the magnetic moment of the young pulsars, it may be recalled that, a typical
 NS or a pulsar has $z <0.25$. Thus  MECOs can act as  extreme  GR pulsars
and hence their role in cosmic particle acceleration and $\gamma$-ray production needs to
investigated. This paper is a highly preliminary study to this effect.

\section{ Unipolar Inductor}
It is known that an isolated spinning NS having a dipole magnetic moment $\mu$ generates
induced {\em vacuum} electric field around it whose components are\cite{14}:
\begin{equation}
E^r = - {\mu \Omega R^2\over c r^4} (3 \cos^2\theta -1); \qquad E^\theta = 
- {\mu \Omega R^2\over c r^4}  (2 \sin\theta \cos\theta)
\end{equation}
Here we have considered an aligned rotator, i.e., one for which the  spin
vector $\vec \Omega \parallel \vec \mu$.
It is this electric field which overwhelms the gravitational pull on ions/electrons
and fling them into acceleration. The parallel electric field at the surface
is
\begin{equation}
E_\parallel \sim {\mu \Omega \over c R^2 } \sim {R \Omega\over c} B_s \sim 2. 10^8 P^{-1} B_{12}
~volts ~cm^{-1}
\end{equation} 
where $B_{12}$ is the value of the surface magnetic field, $B_s$ in units of $10^{12}$ Gauss.
If the accelerating regions would be of the extent of stellar radius, $R \sim 10$ Km,
one would have particle acceleration of $> 10^{14}$ eV even for a slow rotator with
$P = 1$s.
However, despite initial expectations, as of now, there is no direct evidence that
pulsars are sources of UHE cosmic rays. On the other hand, the discovery of MeV-GeV
 $\gamma$-ray pulsars have shown that electrons/pairs are accelerted atlest by a Lorentz
Factor of $ \Gamma \sim 10^{2-3}$ or probably upto $\Gamma \sim 10^{6-7}$\cite{15}. 
The torn off charges along with pairs produced around the initially vacuum NS eventually
form a ``magnetosphere'' which is some sort of a rigid extension of the conducting and spinning
NS. In this
non-vacuum of the magnetosphere, the magnitude of the vacuum electric fields as well
as the associated potential drops reduce significantly. Further, acceleration of charged
particles gets quenched by (a) Curvature Radiation Losses and (b) Inverse Compton Losses.
The energy of the emitted photons can be limited by either (a) $\gamma +\gamma \to e^\pm$
or (b) single photon
pair production, $\gamma + B \to e^\pm +B$, in the strong magnetic field. The break
found in the speectra of the EGRET $\gamma$-ray pulsars could  be because of the latter
process.

As of now no  pulsar is known to emit pulsed VHE $\gamma$-rays and almost all
known $\gamma$-pulsars (probably Geminga too) shine in radio band.
In the following we shall briefly outline the basic difference of spinning MECOs
{vis-a-vis} pulsars and point out why they might be VHE $\gamma$ emitters even being
{\em quiet as radio pulsars}.

\section{GR Unipolar Inductor}
When any massive object rotates, it tries to drag the surrounding spacetime along
with it at a rate $\omega(r)  = J/r^3$ (for slow rotation), where $J = I \Omega$ is the
angular momentum and $I$ is the moment of inertia of the body. The strength of
gravitation of the body is measured by $z$ and recall that while the value of $z \sim 0.2$ 
for a typical NS, the same for a MECO is $z \sim 10^8$. By taking $G=c=1$, the corresponding
induced vacuum electric fields (as seen by a {\em local} observer) is\cite{15}
\begin{equation}
E^r ={\mu (3\cos^2\theta-1)\over 4 M^6 r^3 c}\left\{C[6M^4r^3(2r-3M) \ln N^2+4M^5r(6r^2-3Mr+M^2)]
+ {3\omega M^3r^4\over 2} \ln N^2 + 3M^4\omega r^3\right\} 
\end{equation}

\begin{equation}
E^\theta ={-3 \mu \cos\theta \sin\theta \over  M^6 r^4 N c}\left\{C
[3M^4r^3(r^2 - 3Mr + 2M^2) \ln N^2+2M^5r^2(3r^2-6Mr+M^2)]
+ {\omega M^5r^3\over 2} \right\} 
\end{equation}
where $N^2(r) = (1-r_g/r)$ and $N_R^2 = (1-r_g/R) =(1+z)^{-2}$. The constant $C$ involves
$N_R$, $\Omega$, $M$ and $R$. For extremely high $z \gg 1$,  $R \to r_g = 2M$, and, then,
 we have found that
\begin{equation}
C \sim (2 \ln N_R^2 +3) {\Omega\over r_g} \sim -4 \ln(1+z) {\Omega\over r_g}
\end{equation}
Little away from the MECO surface $r=R$, $E^r$ and $E^\theta$ contain components 
falling off almost as slowly as $\sim \Omega/r$ whereas in a flat spacetime they fall
off as $\sim \Omega/r^4$! For a comparison, at $r= 2R \approx 2r_g$, for a MECO with
$z =10^8$, we have
\begin{equation}
E^r \sim 6000 {\mu \Omega \over c r_g^2 } (3 \cos^2\theta -1)
\end{equation}

Let us compare this with the almost flat NS case with $r = 2R$:
\begin{equation}
E^r = - {\mu \Omega \over 16 c R^2} (3 \cos^2\theta -1)
\end{equation}

Thus, for a $10 M_\odot$ MECO,
 we have the outstanding result that for a given  $\Omega$ and
$\mu$, the basic {\em vaccum accelerating electric field will be higher by a factor
of $\sim 10^4$}. The resultant particle acceleration is not appreciably
quenched by the curvature radiation reaction 
 because curvature losses $\propto \Gamma^{1/4}$.  
A naive scaling would indicate that MECOs thus could the
source of UHE cosmic rays and  $\gamma$-rays.
However, like the pulsar case, the actual available potential drop 
could be much lower because of variety of reasons, yet, we feel, it
should be sufficient to generate UHE cosmic ray acceleration and
VHE $\gamma$-ray production.

\section{Distinct Features of Spinning MECOs}
All $\gamma$-ray  pulsars are likely to be radio pulsars too 
because, (i) Radio emission involves much lower energy pairs
and (ii) the condition of {\em coherent curvature radiation}, believed to
be the reason behind the radio emission, can easily be satisfied in an
almost {\em flat spacetime} around a NS.

But the spacetime around a MECO is completely different: even if the value
of $N(r)^{-1}  = (1- r_g/r)^{-1/2}$ is $\sim 10^8$ at the surface $r= R \approx r_g$,
 its value drops to $\sim\sqrt{ 2.0}$ for $r =2R \approx 2 r_g$! Because of this almost step
function like behaviour of $N(r)^{-1}$ near $r=R$ and further additional distortion
of spacetime by Frame Dragging (rotation), there cannot be any coherent process
occurring either near the surface or any ``Polar Cap''. Thus unless radio signals
originate in some far-off ``outer gaps'' there will not be any pulsed radio emission.

The hot MECO surface however has a highly redsfited quiesent weak  luminosity
of $\sim 10^{30-31}$ erg/s (as measured,  by a distant observer), and this may be 
 UV/Soft X-rays\cite{2,3,4,5}.
And since this originates near the surface, it is unpulsed.

However hard X-ray or $\gamma$ ray emission  is {\em no  coherent} process
and presumably may originate sufficiently away from the surface to become pulsed
with certain distortions.

We have found that because of intense GR effects on the magnetic field
structure, $B$ will be sufficiently low at $r >2R$ to avoid magnetic pair
production of VHE $\gamma$-rays. Thus it might indeed 
be possible  to have (somewhat distorted) pulsed  VHE $\gamma$-ray emission from isolated spinning MECOs. 
But the value of $L_\gamma$ could be less than corresponding pulsar values because of
lower $\Omega$ and $\mu$ except for exceptional cases.
Even if pulsed VHE $\gamma$-ray production would be suppressed by some unforseen
reasons, a spinning MECO  could be generate an outflow of UHE pairs beyond the light cylinder,
i.e., there should be a  relativistic MECO wind. And this MECO wind may generate unpulsed $\gamma$-rays
(nebula) extending far beyond the VHE range.

In fact, there are almost 25 PWNe {\em with no detected pulsar}\cite{16}. Some of these
supposed PWNe could actually be MECOWNe. Interestingly, several of such supposed
PWNes have been detected in $\gamma$-rays by either COMPTEL or EGRET\cite{17}. In other words,
some of the unidentified EGRET sources too can be actually associated with MECOs or
MECOWNe.

\section{Discussion}
We have indicated that isolated spinning MECOs could be a new class of source of VHE $\gamma$-rays
and there may not be any pulsed radio emission associated with them. The (10) Wind
 Nebulae
associated with the  spinning MECOs could 
be yet another class of (unpulsed) VHE $\gamma$-rays. MECOs are born
by the collapse of massive stars and we are certain that the central compact objects
associated with long duration powerful Gamma Ray Bursts harbour MECOs rather than BHs.
Irrespective of this association, the remnant associated with GRBs could be indentified as
VHE $\gamma$-ray emitters. Recall that while Supernova Events release only $10^{48-49}$
erg/s/sr of electromagnetic energy energetic GRBs  emit $10^{51-53}$ erg/s/sr
of electromagnetic energy. Thus GRB remnants should be a class distinct from much more frequent
SNRs. It is likely that the ambient magnetic field in the GRBRs is much higher than in
SNRs. Consequently, (11) GRBRs could be some of the extended ``Dark Accelerators'' because the
central MECO may not generate any pulsed radio emission.

\end{document}